# Outcome-Driven Open Innovation at NASA

**Jennifer L Gustetic**
[a]NASA Headquarters, 300 E St SW, Washington DC, 20546, USA, jennifer.l.gustetic@nasa.gov,

**Jason Crusan**
[a]NASA Headquarters, 300 E St SW, Washington DC, 20546, USA, jason.crusan@nasa.gov

**Steve Rader**
[b]NASA Johnson Space Center, 2101 Nasa Parkway, Houston, TX 77058 USA, steven.n.rader@nasa.gov

**Sam Ortega**
[c]NASA Marshall Space Flight Center, Huntsville, AL 35811, USA, sam.ortega@nasa.gov

Abstract
In an increasingly connected and networked world, the National Aeronautics and Space Administration (NASA) recognizes the value of the public as a strategic partner in addressing some of our most pressing challenges. The agency is working to more effectively harness the expertise, ingenuity, and creativity of individual members of the public by enabling, accelerating, and scaling the use of open innovation approaches including prizes, challenges, and crowdsourcing. As NASA's use of open innovation tools to solve a variety of types of problems and advance of number of outcomes continues to grow, challenge design is also becoming more sophisticated as our expertise and capacity (personnel, platforms, and partners) grows and develops. NASA has recently pivoted from talking about the benefits of challenge-driven approaches, to the outcomes these types of activities yield. Challenge design should be informed by desired outcomes that align with NASA's mission. This paper provides several case studies of NASA open innovation activities and maps the outcomes of those activities to a successful set of outcomes that challenges can help drive alongside traditional tools such as contracts, grants and partnerships.

1. Introduction
In an increasingly connected and networked world, the National Aeronautics and Space Administration (NASA) recognizes the value of the public as a strategic partner in addressing some of our most pressing challenges. The agency is working to more effectively harness the expertise, ingenuity, and creativity of individual members of the public by enabling, accelerating, and scaling the use of open innovation approaches including prizes, challenges, and crowdsourcing. As stated by NASA Deputy Chief Technologist Jim Adams, "NASA recognizes that these methods present an extraordinary opportunity to inspire the development of transformative solutions by offering a means to engage with non-traditional sources of innovative ideas, all in a remarkably cost-effective way" [1].

At NASA, prizes, challenges and crowdsourcing complement our other traditional problem solving approaches to create a robust toolset of innovation approaches for use by a variety of programs. NASA has been a leader in the United States' use of prize competitions for quite some time. The White House recognized this leadership in their 2011 Report to Congress on prize competitions: "From the Centennial Challenges Program, to the NASA Open Innovation

Pavilion, to the NASA Tournament Lab, NASA leads the public sector in the breadth and depth of experience and experimentation with prizes and challenges… [NASA is] best positioned to demonstrate results from the use of prizes and challenges. Examples and case studies from prizes and challenges run by [NASA] highlight[s] what can be expected from all Federal agencies as they begin using prizes for open innovation." [2] Thus NASA is not only seen as a leader in this space, but also as setting the pace for future experimentation and teaching the rest of the government and the world.

NASA is supporting and learning from other Federal agency's prize competitions as well through the NASA Center of Excellence for Collaborative Innovation (CoECI) [3]. The CoECI was launched in November 2011 to advance the use of collaborative innovation techniques to improve Government missions. The CoECI helps NASA centers and other US Federal government programs run their first challenge driven open innovation activities.

This paper will highlight several case studies that show the diversity of purposes and impacts open innovation have in stimulating space-related activities, including:

- Realizing new cost savings and encouraging the development of better products and solutions "on demand"
- Enabling NASA to bring out-of-discipline perspectives to bear and reach beyond the "usual suspects" to increase the number of minds tackling NASA's problem
- Stimulating the development of new commercial markets and thus new opportunities for business and jobs to form

2. Challenge Programs and Definitions

The United States Federal Government has been encouraged to find new and improved ways of solving problems and driving innovation through the use of existing/emerging open innovation tools and challenge platforms. The National Aeronautics and Space Administration (NASA) has adopted policy [4] to encourage the use of challenges, including prize competitions and crowdsourcing activities, to further the Agency's mission at all levels of the NASA organization. This section will describe NASA's definitions for terms such as prize, challenge, and crowdsourcing, which have non standard definitions across sectors and even within the US Federal government. It will also describe the structure and relationships between NASA's various prize and challenge programs.

NASA's Policy Directive 1090.1 [4] defines and explains the terms "challenges", "prize competitions", and "crowdsourcing" as follows. Collectively, these methods are broadly referred to as "Open Innovation" throughout this paper:

- "**Challenges** use a focused problem-statement approach to obtain solutions and/or stimulate innovation from a broad, sometimes undefined, public rather than a specific, named group or individual. Prize competitions and crowdsourcing are two specific techniques for implementing Challenges.
- A challenge implemented as a **prize competition** is intended to stimulate innovation in a manner that has the potential to advance NASA's mission through the offer of a competitive award (e.g., those prize competitions implemented by NASA's Centennial Challenges Program). These challenges are typically administered by NASA or a third party allied organization and offered directly to the public.
- A challenge implemented through **crowdsourcing** is intended to solicit products, services, ideas, or content contributions from many people, often (but not necessarily)

> through the Internet, and may result in the making of award(s) (e.g., NASA Tournament Lab and NASA Innovation Pavilion). An award can be any form of recognition provided to a participant in a challenge, including a cash payment, value other than cash (e.g., payment of travel expenses, accommodation on a launch vehicle) and other forms of reward (e.g., recognition, invitation to an event)." [4] Crowdsourcing may use either NASA employees or external communities, may be for idea generation, strategic technology assay, product or service construct, education/outreach, or may be used to repurpose NASA technologies for earth-space benefit. These challenges are typically run using existing communities that are often organized or "curated" by commercial companies. While the communities are built and maintained by commercial entities, they tend to be open for anyone in the public to join and participate. .

Offices and programs throughout NASA use prizes and crowdsourcing to address challenging problems for the Agency. NASA's Office of the Chief Technologist (OCT) provides strategy and policy for open innovation at NASA and plays a coordinating role for these numerous open innovation programs and projects, including:

- Since 2005, NASA's Centennial Challenge Program [5] has directly engaged the public at large in the process of advanced technology development that is of value to NASA's missions and to the aerospace community through prize competitions. The Centennial Challenges Program is part of the Space Technology Mission Directorate (STMD).
- In 2011, NASA established the CoECI to coordinate NASA's use of challenges implemented through crowdsourcing and to advance the use of open innovation methodologies to improve its own and other Agency missions. The CoECI is a Government-led, virtual center of excellence that serves to harness and redistribute the government's collective experience in, and best practices for, Open Innovation. The CoECI is supported jointly by NASA's Human Exploration and Operations Mission Directorate (HEOMD) and the Office of the Chief Technologist (OCT) and operated with support of the Human Health and Performance Directorate at the Johnson Space Center. The CoECI administers the NASA Tournament Lab [6], the NASA Innovation Pavilion [7], and NASA@Work [8]. CoECI uses these platforms to run challenges for both NASA internal projects and across other government agencies in an effort to introduce and infuse these new open innovation methods across all federal agencies.
- Several NASA offices have supported individual open innovation projects including: the Office of the Chief Information Officer has conducted the Annual International Space Apps Challenge [9] and the Office of Education has supported a number of education challenges including the Robotic Mining Challenge (formerly known as Lunabotics) [10].

3. Moving from Benefits and Types to Outcomes

To date, many of the third party reports and Unites States Federal Government policy and guidance documents on open innovation have focused on the benefits of using the incentive prize approach or structural observations about prize "type". For example, one Office of Management and Budget Memorandum from 2010 lists the following benefits to the prize approach:
- Establish an important goal without having to choose the approach or the team that is most likely to succeed
- Pay only for results
- Highlight excellence in a particular domain of human endeavor to motivate, inspire, and guide others
- Increase the number and diversity of the individuals, organizations, and teams that are

- addressing a particular problem or challenge of national or international significance
- Improve the skills of the participants in the competition
- Stimulate private sector investment that is many times greater than the cash value of the prize
- Further a Federal Agency's Mission by attracting more interest and attention to a defined program, activity, or issue of concern
- Capture the public imagination and change the public's perception of what is possible. [11]

At NASA, we believe that as the sophistication of challenge design grows however, the conversation should not focus on the benefits of the approach or challenge structure itself, but instead on the results and outcomes that challenges can provide. Furthermore, a clear understanding of intended outcomes from a challenge should be a guiding design factor in how challenges are designed and structured from the beginning. Outcomes should not be an afterthought, but instead a guiding factor in the design of any challenge. Challenge design should be informed by desired outcomes that align with NASA's mission.

With nearly 10 years of prize and challenge design experience, NASA challenge managers understand that challenge designs can vary greatly depending on the primary outcomes that a challenge is seeking to further. Based on this experience with nearly 50 challenges, NASA developed an initial listing of the types of outcomes we have seen realized to date through the wide variety of challenge types we have conducted:

1. Research Advancement: The solutions resulting from a challenge identified information the challenge sponsor did not previously know. Solutions enlarged the understanding of the solution space for a particular problem area. Solution space allowed NASA to explore the technical sufficiency of current approaches.
2. NASA Operational Integration/Use: Winning solution(s) was directly integrated into or is being used in NASA operations or a NASA operational environment (e.g. code into a system, testing techniques into a lab, etc...).
3. External Use: The solutions resulting from a challenge are being used by non-NASA entities or individuals (e.g. open source code or other government or commercial buyers).
4. Education/ Public Outreach: The solutions were developed for the purpose of education and are in use by the education community. Alternatively, the solutions from the challenge sought to further outreach or awareness.
5. Advance State of Art/ Demonstrate Proof of Concept: The demonstration of the solution was a proof of concept and/or a game changing demonstration of the possibilities of advanced technology. The purpose was not acquisition of the solution but instead demonstrating the possible and advancing the current state of the art.
6. Enable Product to be Brought to Market: The solutions themselves or closely related spinoffs resulting from the open innovation activity were commercialized by competitors.
7. Creation of New Aerospace Vendors/ Companies: Competitors in the open innovation activity formed into new companies that have become new vendors for NASA and other customers, diversifying the marketplace.

Challenges at NASA have sought to attain one or a number of these outcomes. In the vast majority of the over 50 successful challenges NASA has conducted in the last decade, NASA has experienced at least one of these outcomes. It is important to note that there are still many types of challenges that NASA has not yet conducted, so as NASA's experience with additional challenge types and crowdsourcing platforms grows, the types of outcomes NASA seeks through challenge-driven methods could also grow.

Some case studies from both successes and failures are described in the following section. Learning from the experience from those challenges, we have begun to encourage NASA challenge owners to explicitly identify which of these outcomes they seek at the beginning of challenge design in order to help shape their challenge structure, design, and operational approach. Many early NASA challenges, such as those highlighted as case studies in this paper, were not explicitly designed with this particular outcomes-driven framework in mind. This outcome-driven design approach represents a more mature way NASA is currently designing challenges through its Centennial Challenges Program and CoECI. This maturity in thinking will also assist in performance management across the variety of challenge types conducted across NASA in the future.

Building off NASA and other Federal agency's experience with challenges, Deloitte University Press released a report in 2014 that looked through over 400 challenges that have been conducted since 2009 and categorized the results the U.S. Government is seeing into six general outcomes. The seven outcomes listed previously in this section are specific to NASA's missions and the types of challenges we have run to date. The Deloitte report looks more generally at prize outcomes for the entire federal government and philanthropic community and identifies six outcomes that designers commonly seek (individually or in combination), falling along two dimensions.

- Developing ideas, technologies, products, or services
    - Attract new ideas
    - Build prototypes and launch pilots
    - Stimulate markets
- Engaging people, organizations, and communities
    - Raise awareness
    - Mobilize action
    - Inspire transformation [12]

There are many parallels between NASA's working list of seven outcomes and the six identified in the Deloitte report and more work needs to be done to tailor outcomes to specific agency missions to assist in agency level performance measurement and management. The remainder of this paper presents some case studies, mapped to working list of seven NASA specific outcomes related to open innovation activities.

4. Case Studies

This section provides six case studies of recent incentive prizes and crowdsourcing activities. These case studies were presented at the 65th International Astronautical Congress. In this paper, these case studies are mapped more clearly to the outcomes list NASA has been developing. These case studies illustrate the wide variety of results generated through challenges as well as the numerous outcomes they can help drive. Note this table maps the actual results of the case studies to date to the proposed outcomes framework. Table 1 provides a high level view of the each case study and their related outcomes.

| Challenge | Research Advancement | NASA Operational Integration/ Use | External Use | Education/ Outreach | Advance State of Art/ Demonstrate Proof of Concept | Enable Product to be Brought to Market | Creation of New Aerospace Vendors/ Companies |
|---|---|---|---|---|---|---|---|
| | | | | | | | |
| Astronaut Glove Challenge | | | X | | X | X | X |
| Lunar Lander Challenge | X | X | | | X | X | X |
| International Space Station (ISS) Longeron Shadowing Challenge | | X | | X | | | |
| Kevlar and Vectran Strain Measurement Challenge | | X | | | | | |
| Non-Invasive Intracranial Pressure Measurement Challenge | X | | | | | | |

**Table 1: Selected Case Studies and Outcomes**

One of the case studies described below is not present in Table 1 because it was an unsuccessful challenge. This challenge is presented last in the sequence of case studies to demonstrate that in addition to many successes their have been a few failures in space open innovation activities that NASA has learned from in designing future prize, challenge and crowdsourcing activities.

Many early NASA challenges, such as those highlighted in the following case studies, were not explicitly designed with this particular outcomes-driven framework in mind. Alternatively, the results from these early challenges have informed the outcomes-driven framework that NASA challenge designers are currently using in the design of current challenges while learning from the experience of previous ones. This outcome-driven design approach represents a more mature way NASA is currently designing challenges through its Centennial Challenges Program and CoECI.

4.1 Astronaut Glove Challenge

The Astronaut Glove Challenge was a dual level competition conducted by NASA's Centennial Challenges Program from 2007-2009 seeking improvements to glove design that would reduce the effort needed to perform tasks in space and improve the durability of the glove.

In the pressure suits that astronauts must wear while performing a spacewalk, one of the toughest parts to design are the gloves. Like an inflated balloon, the fingers of the gloves resist the effort to bend them. Astronauts must fight that pressure with every movement of their hand, which is exhausting and sometimes results in injury. Furthermore, the joints of the glove are subject to wear that can lead to life-threatening leaks. In this challenge, competitors demonstrated their glove design by performing a range of tasks with the glove in an evacuated chamber similar to use in space and to test the gloves for leaks. In order to qualify for a prize, the gloves had to meet all of the basic requirements and also exceed the flexibility of the current NASA spacesuit glove. In the 2007 competition, only the pressure-restraining layer was required. For the 2009 Challenge, teams had to provide a complete glove, including the outer, thermal-micrometeoroid-protection layer that protects the pressure-restraining layer from the environments of space and the inner, pressure-restraining layer.

As with most of NASA's Centennial Challenges a non-reimbursable (unfunded) agreement was signed with a non-profit organization to conduct the challenge. Volanz Aerospace Inc., a non-profit space education organization based in Owings, Maryland was selected to be the allied organization. They secured a commercial sponsor for the event, Secor Strategies LLC of Titusville, to help cover the costs of the event.

The 2007 competition for the demonstration of a better inner glove the prize purse was set at $250K. With five teams registered to compete three teams showed up with gloves to test. The 2007 winner was Peter Homer from Southwest Harbor, Maine. An unemployed aerospace engineer at the time, Peter used the prize money to start a space glove company Flagsuit LLC who now provides gloves to NASA spacesuit vendors. The 2009 competition had two teams show up to compete demonstrating the complete spacewalk glove; Peter Homer the previous winner and Ted Southern who competed in 2007 and partnered with Nik Moiseev, a member from the third team registered in the 2007 competition. The interesting thing to note is the partnering that occurred between 2007 and 2009. With Nik and Ted joining together they were able to develop a glove to win second place in 2009 for a prize purse of $100K with Peter just barely winning first place with a prize purse of $250K. Nik and Ted have formed Final Frontier Design, a full spacesuit company supplying suits to a company in Spain and to Starfighters Aerospace operating out of Kennedy Space Center.

4.2 Lunar Lander Challenge

The Lunar Lander Challenge conducted by NASA's Centennial Challenges Program from 2006-2009 involved building and flying a rocket-powered vehicle that simulated the flight of a vehicle on the Moon. The lander had to take off vertically then travel horizontally and land accurately at another spot. Then the same vehicle had to take off again, travel horizontally back to the original take off point and land successfully.

The Northrop Grumman Lunar Lander X PRIZE Challenge was a $2M incentive prize program designed to build an industry of American companies capable of routinely and safely flying vertical take-off and landing rocket vehicles useful both for lunar exploration and for other applications. This prize, along with the Google Lunar X Prize is part of an effort to jumpstart "Moon 2.0," a new era of sustainable lunar exploration that involves international partnerships between government space agencies and entrepreneurial firms.

The XPrize Foundation through a non-reimbursable (unfunded) agreement with NASA conducted the challenge. The prize purse came from NASA's Centennial Challenges program, with majority of the operational funding coming from Northrop Grumman. The prize was divided into two levels—a relatively easier Level One and a more difficult Level Two—each of which has a first and second place prize; $350K and $150K for Level One and $1M and $500K for Level Two. Level One required flight duration of at least 90 seconds on each flight and Level Two required flight duration of at least 180 seconds on each flight. Furthermore, one of the landings for a Level Two attempt had to be made on a simulated lunar terrain with rocks and craters.

The prize was announced in May 2006, and 2006, 2007 and 2008 was offered to teams only on a fixed date in late October and in a fixed location (Southern New Mexico, with the generous support of Spaceport America). Eleven teams competed throughout the 4 years of the competition. At the conclusion of the 2008, Armadillo Aerospace, led by id software founder John Carmak, claimed the $350,000 Level One, First Place prize. In 2009, the competition was altered to allow teams to compete at a date and a location of their choosing at any point between

early August and the end of October. At the end of that period, teams that had met all of the requirements for the prize would be ranked according to the landing accuracy displayed on their two flights. Masten Space Systems won first place and Armadillo Aerospace won the second place prize. Both teams have since received contracts from NASA to develop vehicles to provide suborbital flight opportunities to conduct microgravity science.

Additionally, Masten Space Systems has entered into a competitive selected no-funds-exchanged Space Act Agreement (SAA) partnership with NASA in April 2014 to spur commercial cargo transportation capabilities to the surface of the moon. NASA's Lunar Cargo Transportation and Landing by Soft Touchdown (Lunar CATALYST) initiative is establishing multiple no-funds-exchanged SAA partnerships with U.S. private sector entities. The companies, Astrobotic Technologies of Pittsburgh, Pa., Masten Space Systems Inc. of Mojave, Calif. and Moon Express Inc., of Moffett Field, Calif., will not only develop capabilities that could lead to a commercial robotic spacecraft landing on the moon but also potentially enable new science and exploration missions of interest to NASA and to broader scientific and academic communities. It is important to note all three teams that won this competitive partnership opportunity are prize teams, having participated in either the lunar lander challenge (Masten) or the Google Lunar X Prize (Astrobotic and Moon Express).

4.3 International Space Station (ISS) Longeron Shadowing Challenge

The challenge conducted by the Center of Excellence for Collaborative Innovation (COECI) in 2012 was to develop a new algorithm that would show how to position the solar arrays on the ISS to generate as much power as possible during the most difficult orbital positions. Additionally, this challenge included an outreach component; a T-Shirt design contest to promote the main challenge.

While pointing the solar arrays directly at the Sun is fairly straight forward, the ISS must also consider the loads and stresses on the longerons which support the arrays which can be caused by shadowing due to the ISS modules and structure. If these support structure materials are exposed to extreme heat from the Sun on one section and the cold of deep space on an adjacent section, this can cause expansion and contraction of the materials thus causing a stress that damage the structure. This algorithm is needed to ensure that the solar arrays can be pointed to get the most energy while avoiding or mitigating conditions where these shadows can occur. Power is the key to successful ISS operations and research and addressing this issue serves to both enable the operations and research and prolonging the service life of the solar array systems.

This challenge to develop a new algorithm (ISS already has one it uses operationally) was lead by William Spetch in the ISS Program at Johnson Space Center in Houston, Texas. Bill worked with NASA's COECI to formulate a challenge on the NASA Tournament Lab (NTL) which could be run on the TopCoder platform under a NASA contract with Harvard University. Spetch and his team spent several months formulating the problems so that the full complexity of the issue was described and adapting simulations that the challengers could use to test their algorithms. In January of 2012, TopCoder launched a 3 week challenge to develop a new algorithm to see if we might do better with a different approach than the current algorithm. The challenge paid out over $40,000 in prizes and attracted over 4000 registered solvers. In the end, 459 competitors made 2185 submissions from which 10 winners were selected. This number of competitors broke a new record for the NTL.

The resulting algorithms were shown to be as good as the ISS current tools for core operations and a significant improvement for many of the edge cases. There is some candidate follow on work being considered, but it mostly helped to inform possible improvements to the existing systems & processes ISS uses for this function.  The ISS mainly used this as a pilot to understand challenges and thus did not have a concrete plan or path to take the results into full operations.  It has been demonstrated in multiple challenges that unless the challenge owner has plans, resources, and authority/buy-in to operationally implement the solution from a challenge, it will likely not be integrated into operational systems, regardless of the quality of the resulting solution.

Funding for this challenge was provided from NASA's HEOMD Strategic Operations Budget.  The total cost of the challenge including awards and operational expenses was $109,600.  NASA estimates that a similar internal effort would have cost over $240,000.

4.4 Kevlar and Vectran Strain Measurement Challenge

This challenge conducted by the COECI in 2012-2013 was seeking a new method to measure the strain on Kevlar and Vectran straps in the 25 to 125°C range. Measurement by traditional contact extensometers has caused damage and premature failure while non-contact methods such as photogrammetry have worked well with certain samples and at room temperature, but some samples, where the fibers twist and bulge during the measurement at the elevated temperature range, cannot be measured. A technique was needed to accurately measure the strain in these samples, given the fiber movement.

This challenge is important to NASA's space exploration because it is a key enabler to new lightweight, inflatable structures.  The woven materials used to build these structures have a tendency to stretch under load and so NASA must be able to test and characterize these materials under various environments.  The Challenge is to find a method to measure strain on the Kevlar and Vectran straps with high differential fiber movement at temperatures from 25 to 125C that performs as well as photogrammetry and eliminates the problems where the fibers twist and bulge as with current techniques. 

This challenge was lead out of NASA's Langley Research Center (LaRC) and their OCT/GCTD/Lightweight Materials & Structures group by Melvin Ferebee, Lynn Bowman, and Tom Jones and was funded by Langley's Center Director Discretionary Funds.  The team worked with Innocentive to post a challenge on October 25, 2012, running through January 2, 2013 with $20,000 in available award money.   This challenge attracted 347 solvers from 45 different countries around the world.  There were 71 solutions submitted from 19 different countries and 26 of those solutions were considered.   In the final evaluation, the team selected 3 winners (awarded $10K, $5K, and $5K).  Two solutions were from the United States and one solution from Serbia. The winning solutions suggested adding a strip of elastomeric material (a rubber strap) along with the woven strap in the test jig.  The measurements are then taken off of the rubber strap and correlated back to the woven strap.

The team considered these solutions to be immediately useful by the analysis team and applicable to multiple projects outside of Lightweight Materials and Structures.  The team considered this solution "extremely elegant, simple and repeatable-- it took a fresh perspective to cultivate a potential gem". By approaching this problem through open innovation, NASA saved the taxpayer dollars by not contracting out a lengthy research program to seek an answer.

4.5 Non-Invasive Intracranial Pressure Measurement Challenge

This challenge conducted by the COECI in 2012-2013 was to find a non-invasive method to measure intracranial pressure. During spaceflight, the astronaut's body experiences short and long term changes in physiology which may result in permanent changes to tissues and organs, especially during long missions. NASA has documented that some astronauts who have been on long duration missions (6 months in microgravity) experience changes in visual acuity and in eye anatomy. NASA suspects that these changes in the eye are related to increased intracranial pressure and would like to monitor this pressure non-invasively over time. Currently, known measurement technologies are either invasive (lumbar puncture or cranial implant) or too inaccurate to be acceptable for repeated measurements over time. which are both too invasive. Coming into this challenge, NASA did not have any proven method to quantify intracranial pressure *non-invasively* and techwatch and market surveys indicated that the current state of technology was insufficient to meet NASA's research needs.

This challenge was sponsored by NASA's Human Research Program and lead by Jennifer Villarreal. The team decided to solve this problem using multiple approaches. The team initially ran a challenge on NASA's internal challenge platform, NASA@work which poses the problem to the NASA employee and contractor population. This challenge resulted in 3 NASA@work winners whose inputs directed the team to take a second look at developers that the team had previously eliminated from consideration. The team then ran the challenge on both the Innocentive and Yet2.com platforms.

The Innocentive challenge was posed as a theoretical challenge with a prize of $15,000. The challenge posted on December 17, 2012 and resulted in 636 submissions. Through the evaluation process, 46 solutions ended up in the final review and two solutions were selected as winners. One winner was UCLA's Non-invasive Intracranial Pressure Calibration Framework (NICF). This was an algorithm developed (under an NIH grant) from a database of cerebro-vascular parameters, non-invasive and invasive ICP measurements. It also used the cerebral blood flow velocity signal measured at the middle cerebral artery using conventional Transcranial Doppler (TCD) ultrasound and arterial blood pressure signals. This was particularly advantageous method since the TCD measurement is available with existing Ultrasound on the International Space Station (ISS). The other winner was called "Thinker: An Intelligent Intracranial Pressure Monitor." This was a miniature physiological data acquisition system with an algorithm that predicts ICP from digitized pressure waves, i.e. plethysmography. This was based on a commercial digital technology with innovative sensor architecture, electronics and software/firmware techniques. While this technology showed significant promise, the algorithm would require improvements.

Additionally, the team posted the challenge on Yet2.com which is a tech scouting firm. This challenge resulted in 81 identified leads of which 3 were solutions of high interest. The challenge also resulted in finding 6 other interesting solutions and 6 potential complementary technologies. One of the high interest solutions was NeuroDx which uses the carotid artery pulse pressure waveform that contains signals that are informative of the compliance in the cerebral vessels and thus, ICP. Another high interest solutions was Remote Vital Monitoring which was a device that uses a surface electrode and an auditory stimulation device, attached to the patient's ears, to non-invasively capture and monitor intracranial pressure using the Brainstem Auditory Evoked Response (BAER).

Overall, the resulting solutions assisted the team in better understanding the full breadth and depth of the available technologies and help them to map out a much clearer path to a final usable, mature implementation.

4.6 Strong Tether Challenge

The Challenge conducted by NASA's Centennial Challenges Program from 2005-2011 was to develop a material that is both strong enough and light enough to support a 60,000 mile long tether. Compared to the best commercially available tether, that would be a material that is almost 25 times better - about as great a leap as from wood to metal.

The concept of a space elevator to gain access to low earth orbit has intrigued space explorers since it was proposed by Konstantin Tsiolkovsky in1895. Its main component is a ribbon-like cable (also called a tether) anchored to the surface and extending into space. It is designed to permit vehicle transport along the cable from a planetary surface, such as the Earth's, directly into space or orbit, without the use of large rockets.  The single most difficult task in building a Space Elevator is achieving the required tether strength-to-weight ratio. The Strong Tether Challenge sought to drive material science technologies to create long, very strong cables with the exceptionally high strength-to-weight ratio. Such tethers would enable advances in aerospace capabilities including reduction in rocket mass, habitable space structures, tether-based propulsion systems, solar sails, and even space elevators. Dramatically stronger and lighter materials would also revolutionize the engineering of down-to-earth structures such as aircraft bodies, sporting good equipment, and even structures of bridges and buildings.

The Centennial Challenges Program signed a non-reimbursable (unfunded) agreement with the non-profit organization Spaceward Foundation in Mountain View, CA to conduct the challenge. They conducted the challenge at the Space Elevator Conference from 2005 through 2011 offering a $2M prize for the development of a tether that met the challenge rules for length, weight, and strength. It was assumed that Carbon Nanotube (CNT) material would be the material of choice to develop a capable tether. With over ten teams competing from 2005 through 2011, no one was able to develop a tether with a strength to weight ratio better than the current commercial products available.  After five years of conducting the challenge it was cancelled due to lack of technology improvement.

A post challenge assessment provided the lesson that an incentive prize may not be the best tool for a technology development and demonstration/proof-of-concept challenge when fundamental research and development is still required to advance technology. For this technology area, it would be more feasible to use a challenge to integrate preexisting technologies and techniques in unique ways to create disruptive and innovate technologies to solve the challenge rather than trying to develop and demonstrate new and unknown technologies. While challenges have been shown to advance research in a number of areas, sometimes the competitors required to advance research are different than those required to do systems engineering to combine and demonstrate technologies for new capabilities. Thus, the Centennial Challenge Program, as a program that seeks to demonstrate technologies, has selected challenge topics that require less fundamental research and development in the creation of solutions than was required for the Strong Tether Challenge.

5 Conclusions

NASA's use of open innovation tools to solve a variety of types of problems and advance of number of outcomes continues to grow. Challenge design is also becoming more sophisticated as our expertise and capacity (personnel, platforms, and partners) grows and develops.

As understanding of how open innovation approaches can advance scientific discovery and technology development grows, more study should be focused on performance measurement and Pre-Publication Draft to Space Policy, DOI: 10.1016/j.spacepol.2015.06.002management of these approaches. Since the types, outputs and outcomes of challenges vary so greatly, developing a set of metrics to guide performance management and resource allocation for these types of activities is a difficult, but important task.

Also, additional study should look at how to appropriately identify problems that are well suited to open innovation approaches. Problem identification, definition, and decomposition are critically important steps in running and open innovation activity and can be the most difficult steps. Understanding how program and project managers can identify the problems within their overall program and project that could benefit from these approaches is paramount to making open innovation a tool more regularly used in the program manager's toolkit.

6 Acknowledgements

The following NASA leaders are acknowledged for being critical champions for open innovation work at NASA: Lori Garver, Robert Lightfoot, Rebecca Spyke-Keiser, Bobby Braun, Mason Peck, David Miller, Joe Parrish, W James Adams, Waleed Abdalati, Ellen Stofan, Michael Gazarick, James Reuther, Deborah Diaz, Jaiwon Shin, William Gerstenmaier, John Grunsfeld, Jeri Buchholz, Leland Melvin, Pete Worden, Ellen Ochoa, Jason Crusan, Doug Comstock, Jeff Davis.

Many people have managed or supported open innovation or prize programs at NASA since 2005. The breadth of open innovation work conducted at NASA to date would not have been possible without the contributions from the following current or former NASA employees and support contractors: Ken Davidian, Larry Cooper, Eric Eberly, Janet Sudnik, Andy Petro, Jenn Gustetic, Gladys Henderson, Sam Ortega, Lynn Buquo, Steve Rader, Carolyn Woolverton, Allison Wolff, Kathryn Keeton, Karl Becker, Mike Ching, Carol Galica, Mike Hetle, Beth Beck, Amy Kaminski, Nicolas Skytland, Ali Llydwelin, Sean Herron, Chris Gerty, Diane Detroye, Gloria Murphy, Julie Clift, Deidre Williams, Jared Crooks, Karen James, Jeff Heninger, Randall Suratt.,

Engineers, scientists and technologists that have sponsored or led specific challenges described in the case studies in this paper: Melvin Ferebee, Jennifer Villereal, William Spetch, Thomas Jones, Gabriel Udomkesmalee, James Cockrell, David Korsmeyer, and many more.7 References
[1] Implementation of Federal Prize Authority: Fiscal Year 2013 Progress Report. [Internet]. 2014. p.8. [cited 2015 Feb 15]. Available from: http://www.whitehouse.gov/sites/default/files/microsites/ostp/competes_prizesreport_fy13_final.pdf
[2] Center of Excellence for Collaborative Innovation [Internet]. Washington DC: National Aeronautics and Space Administration.; c2011-15 [updated 2015 Feb 11; cited 2015 Feb 15]. Available from: http://www.nasa.gov/offices/COECI/index.html.
[3] Implementation of Federal Prize Authority: Progress Report. [Internet]. 2012. p.14. [cited 2015 Feb 15]. Available from: http://www.nasa.gov/sites/default/files/files/competes_report_on_prizes_final.pdf.
[4] NASA Policy Directive 1090.1 [Internet]. [updated 2014 Feb 12; cited 2015 Feb 15]. Available from: http://nodis3.gsfc.nasa.gov/displayDir.cfm?t=NPD&c=1090&s=1.
[5] Centennial Challenges Program [Internet]. Washington DC: National Aeronautics and Space Administration; c2005-15 [updated 2015 Feb 12; cited 2015 Feb 15]. Available from: http://www.nasa.gov/challenges.12